\documentclass[conference]{IEEEtran}
\IEEEoverridecommandlockouts
\usepackage{cite}
\usepackage{amsmath,amssymb,amsfonts}
\usepackage{algorithm}  
\usepackage{algpseudocode}
\usepackage{graphicx}
\usepackage{textcomp}
\usepackage{xcolor}
\usepackage{verbatim}
\usepackage{tikz-cd}
\def\BibTeX{{\rm B\kern-.05em{\sc i\kern-.025em b}\kern-.08em
    T\kern-.1667em\lower.7ex\hbox{E}\kern-.125emX}}

\makeatletter
\newcommand{\linebreakand}{%
  \end{@IEEEauthorhalign}
  \hfill\mbox{}\par
  \mbox{}\hfill\begin{@IEEEauthorhalign}
}
\makeatother

\definecolor{GreenColor}{rgb}{0.137,0.5,0.3}

\begin{document}
\title{For One-Shot Decoding:\\Self-supervised Deep-learning Polar Decoder}

\author{\IEEEauthorblockN{Huiying Song}
\IEEEauthorblockA{
\textit{Tokyo Institute of Technology}\\
Tokyo, Japan \\
song.h.aa@m.titech.ac.jp}
\and
\IEEEauthorblockN{Yihao Luo}
\IEEEauthorblockA{
\textit{Imperial College}\\
London, UK \\
y.luo23@imperial.ac.uk}
\and
\IEEEauthorblockN{Yuma Fukuzawa}
\IEEEauthorblockA{
\textit{NBL corporation Japan}\\
Tokyo, Japan \\
yumafukuzawa.CT@gmail.com}

}

\maketitle

\begin{abstract}
We propose a self-supervised deep learning-based decoding scheme that enables one-shot decoding of polar codes. In the proposed scheme, rather than using the information bit vectors as labels for training the neural network (NN) through supervised learning as the conventional scheme did, the NN is trained to function as a bounded distance decoder by leveraging the generator matrix of polar codes through self-supervised learning. This approach eliminates the reliance on predefined labels, empowering the potential to train directly on the actual data within communication systems and thereby enhancing the applicability. Furthermore, computer simulations demonstrate that (i) the bit error rate (BER) and block error rate (BLER) performances of the proposed scheme can approach those of the maximum a posteriori (MAP) decoder for very short packets and (ii) the proposed NN decoder (NND) exhibits much superior generalization ability compared to the conventional one. 
\end{abstract}


\section{Introduction}
Polar codes have garnered significant attention as the control channel coding scheme in the fifth-generation (5G) standard, owing to their capacity-achieving property and low encoding and decoding complexities. Over the past decade, substantial efforts have been dedicated to optimizing the decoding of polar codes, focusing on enhancing performance, reducing complexity, and minimizing latency. For instance, the successive cancellation list (SCL) decoder \cite{b1}, employed in 5G wireless communications, demonstrates near-optimal performance, and the min-sum algorithm (MA) effectively reduces the complexity of belief propagation (BP) decoder \cite{b2}. Concurrently, the rapid advancements in deep learning techniques have revolutionized diverse fields, encompassing computer vision, natural language processing, and speech recognition. Consequently, the potential of combining deep learning with channel coding technologies has sparked increasing interest, prompting extensive endeavors to optimize the decoding of polar codes using deep learning techniques in recent years.
\subsection{Related Work}
Previous studies on leveraging deep learning techniques in the decoding of polar codes have utilized neural network (NN) learning to optimize the conventional decoders. The first example of using deep learning to train the weights assigned to the edges of the Tanner graph, leading to improved performance of the belief propagation (BP) decoder, is introduced in \cite{b3}. Subsequently, in \cite{b4}, the approach of employing deep learning to train the scaling parameters for the scaled min-sum BP decoder of polar codes \cite{b5} is proposed.
The idea of letting the NN learn to decode, enabling one-shot decoding of polar codes, was generated in the pioneering work by Gruber et al. \cite{b6}. In their study, supervised deep learning is employed to train the NN decoder (NND), with noisy words received from additive white Gaussian noise (AWGN) channels as input data, while the correct information bit vectors are used as labels. Their findings demonstrate the ability of the NND to generalize for structured codes, indicating that NNs can learn a decoding algorithm rather than solely serving as classifiers. Afterwards, studies focusing on the integration of one-shot NN decoding with successive cancellation (SC) \cite{b7} and BP \cite{b8} decoders have emerged, aiming to achieve efficient decoding of polar codes with long blocklengths.
\subsection{Motivation of the Paper}
The reliable transmission of short packets has become an increasingly critical requirement in the context of 5G and beyond, particularly due to the emergence of novel traffic types such as machine-to-machine (M2M) communications \cite{b9}, which heavily rely on short packet transmissions. Nevertheless, the state-of-the-art decoding schemes grounded in classic information theory fail to perform well for short packets, since the law of large number cannot be put to work in this context \cite{b10}. In contrast, it is proved in \cite{b6} and this paper that the bit error rate (BER) and block error rate (BLER) performances of the NNDs can approach those of the maximum a posteriori (MAP) decoder for (16, 8) polar codes. Moreover, the current decoding schemes of polar codes face challenges in meeting the stringent low-latency requirements of 5G and future communication systems. Therefore, in this scenario, the NND may be a promising alternative due to its one-shot decoding capability. \\
\indent Although the supervised deep learning-based decoding scheme in \cite{b6} demonstrates near-optimal performance for very short packets, it comes at the cost of training with the entire codebook, i.e., all the different information bit vectors should serve as labels, which is impractical even for short packets in communication systems. Additionally, the decoder exhibits limited generalization ability, making it impossible to achieve good performance using only a subset of the codebook for training. As a result, for instance, the decoder cannot be used for decoding (128, 64) polar codes until it finishes the training with the $2^{64}$ different codewords.\\
\indent To address the aforementioned challenges and devise a suitable NND for the transmission of short packets with low latency, we propose a self-supervised deep learning-based decoding scheme. Specifically, by inheriting the spirit of the bounded distance decoder, we train the NND to output the decoded bit vector, which can be re-encoded to the codeword that exhibits the closest distance with the received word from AWGN channels, i.e., the input data to the NN. This method enables effective recovery of the information bit vector, provided that the number of error bits incurred by the channel is smaller than half of the minimum Hamming distance of the codewords, which is consistent with the high reliability requirement in the communication systems. Computer simulations demonstrate that the proposed NND achieves a significantly improved generalization ability compared to the conventional one presented in \cite{b6}. Besides, in contrast to a predetermined signal to noise ratio (SNR) used for training in \cite{b6}, our experiments employ randomly generated SNRs during the training process, which offers the potential to train the NND using actual received data encountered in communication systems.

\section{Polar Codes}
Polar codes, introduced by Arikan in 2009 \cite{b11}, exploit the channel polarization effect generated by recursive construction approach to systematically transform a set of noisy channels in to subsets of reliable and unreliable channels. By transmitting information bits through reliable channels and frozen (redundant) bits through unreliable channels, polar codes can achieve channel capacity.
\subsection{Encoding}
The encoding of polar codes is formulated by
\begin{equation}
    x = u\mathbf{G}_{N} = u\mathbf{G}^{\otimes n}=u\begin{bmatrix}
1 & 0\\
1 &1 
\end{bmatrix}^{\otimes n},
\end{equation}
where $x$ is the encoded vector, $u$ is the original vector (including information and frozen bits), $N := 2^n$ is the blocklength for $n>0 \in \mathbb{Z}$, and $\otimes$ denotes the Kronecker product \cite{b12}. The Kronecker power is defined inductively by $\mathbf{G}^{\otimes n} = \mathbf{G} \otimes \mathbf{G}^{\otimes (n-1)}$ for positive integer $n$ with $\mathbf{G}^{\otimes 0}$ set to scalar $1$.

\indent Define a $K$-index set $\mathcal{A}$ from $\{1,...,N\}$ as the information index set. The sub-vector $u_{\mathcal{A}} \in \mathbb{Z}_2^K$ formed by $u_i \in \mathbb{Z}_2$ for $i \in \mathcal{A}$ defines the information bit vector, while $u_j \equiv 0$ for $ j \notin \mathcal{A}$ is defined as the frozen (redundant) bit which forms the sub-vector $u_{\mathcal{A}^c} = (0,\cdots,0) \in \mathbb{Z}_2^{N-K}$. 
Then  $u_{\mathcal{A}}$ and $u_{\mathcal{A}^c}$ are combined as the original code $u \in \mathbb{Z}_2^N$. $K/N$ is denoted as the code rate $R$ of $u$. Correspondingly, the generator matrix of polar codes is defined as $\mathbf{G}_{N}(\mathcal{A})$, which denotes the sub-matrix of $\mathbf{G}_{N}$ formed by the rows with indices in $\mathcal{A}$. Therefore, equivalently, the encoding of polar codes can be expressed as $x=u_{\mathcal{A}}\mathbf{G}_{N}(\mathcal{A})$. In this paper, we adhere to the selection of $\mathcal{A}$ as specified in the 5G standard documentation.

\subsection{Non-deep-learning Decoders}
The SC and BP are prominent decoding schemes for polar codes. The SC decoder utilizes a recursive approach, enabling decoding complexity of $O(N\log N)$ for a blocklength of $N$ \cite{b11}. Nevertheless, the SC decoder necessitates a specific decoding order. This lack of parallelization results in high latency, which poses a challenge for its applications in real-time cases.

\indent On the other hand, the BP decoder employs a probabilistic decoding approach based on the principles of message passing in graphical models. It constructs a Tanner graph representation of polar codes and performs iterative message passing between the variable nodes and check nodes \cite{b2}. While BP exhibits a more parallelizable structure compared to SC, the latency remains non-negligible as the number of iterations increases. 


\section{Self-supervised Neural Polar Decoder}

\subsection{NN and Self-supervised Deep Learning}
NNs are computational models composed of interconnected artificial neurons organized into layers. Each neuron receives input signals, performs a weighted sum of these inputs, applies an activation function and produces an output. During training, the network adjusts its weights using optimization algorithms such as backpropagation \cite{b13}. Backpropagation calculates the gradients of a loss function that quantifies the difference between the network's predicted output and the expected output. These gradients are then used to update the weights, iteratively improving the network's predictions. \\
\indent Self-supervised deep learning is a paradigm within the field of unsupervised learning. Unlike traditional supervised learning, which relies on labeled data for training, self-supervised learning leverages the inherent structure and information within the data itself. By formulating pretext tasks, self-supervised learning creates supervisory signals for training deep NNs. Through solving these tasks, the model learns to capture meaningful patterns and relationships from the unlabeled data. \\
\indent Since polar codes have the intrinsic structure described by the generator matrix, self-supervised learning is suitable for training the NND of polar codes by exploiting the generator matrix.

\subsection{Minimum Hamming Distance Principle}\label{subset:MHamming}
A metric of a code $\mathcal{C}$, defining the distance $d(\cdot,\cdot)$ between codewords, involves the minimum distance $d(\mathcal{C})$ of the code as 
\begin{equation}
    d(\mathcal{C}):=\min \{d(u,v):\forall u,v \in \mathcal{C}, u \neq v\}.
\end{equation}
A large minimum distance guarantees the distinguishability of codewords in a code. Therefore, in the classic coding area, the design principle endows codes with a strong algebraic structure and a large enough minimum distance \cite{b14}. When a codeword is transmitted through a channel with limited power of the noise, the noisy word can be bounded within a ball (the set of words with the distance at most the radius of the ball from the center) centered at the transmitted codeword. If the balls of all the codewords are disjoint, the recovery of the transmitted codeword can be established by searching the center of the ball where the noisy word locates.

\begin{figure}
    \centering
    \includegraphics[width=0.45\textwidth]{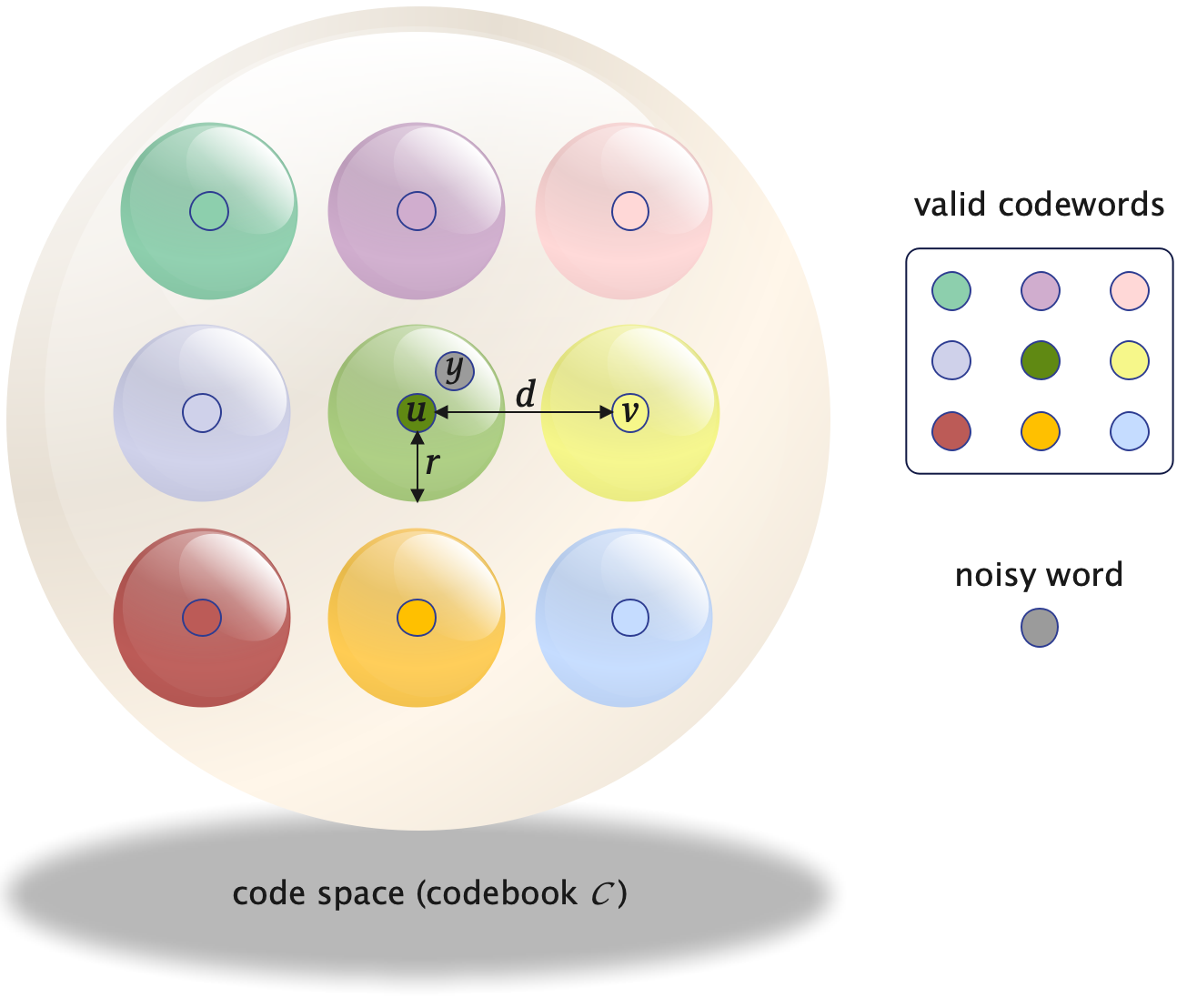}
    \caption{Illustration of Hamming ball}
    \label{fig:HammingBall}
\end{figure}

In the case that $\mathcal{C}$ is a binary code, the Hamming metric $h(\cdot,\cdot)$ is commonly used and defined as

\begin{equation}
    h(u,v) := \sum_{i=1}^N u_i \oplus v_i, 
\end{equation}
where $u,v \in \mathbb{Z}_2^N$, $u_{i}$ and $v_{i}$ are the $i$-th bits of $u$ and $v$, respectively, $N$ is the blocklength, and $\oplus$ denotes exclusive OR. As depicted in Fig. \ref{fig:HammingBall}, a Hamming ball of radius $r\in \mathbb{N}$ centered at the codeword $u$ is defined as $B_r(u):=\{a \in \mathbb{Z}_2^N: h(u,a)\leq r\}$.  It is obvious that for the code $\mathcal{C}$ with minimum Hamming distance $h(\mathcal{C})$, all the Hamming balls of radii less than or equal to $r = \lfloor \frac{h(\mathcal{C})-1}{2} \rfloor$ surround the codewords are disjoint. 

 During the decoding process, given a received word $y$ from a binary symmetric channel (BSC), the so-called bounded distance decoder with searching radius $r = \lfloor \frac{h(\mathcal{C})-1}{2} \rfloor$ outputs $\hat{u}(y)$ by the following rule
\begin{eqnarray}
		\hat{u}(y)= \left\{
		\begin{aligned}
			&u \in \mathcal{C},\quad  \mathrm{if} \ h(u,y) \leq r,\\
			&\mathrm{error},\  \quad \mathrm{if\ no\ such}\ u \ \mathrm{exists},
		\end{aligned}
		\right.
\end{eqnarray}
due to the unique solution for $h(u,y) \leq r$. Thus, the decoder can recover the the correct transmitted codeword, with the number of error bits no more than $r$.

The purpose of this paper is to establish a deep-learning decoder conforming to the bounded distance decoding principle. 

\subsection{Self-supervised Network Architecture}
\begin{figure*}
    \centering
    \includegraphics[width=1.0\textwidth]{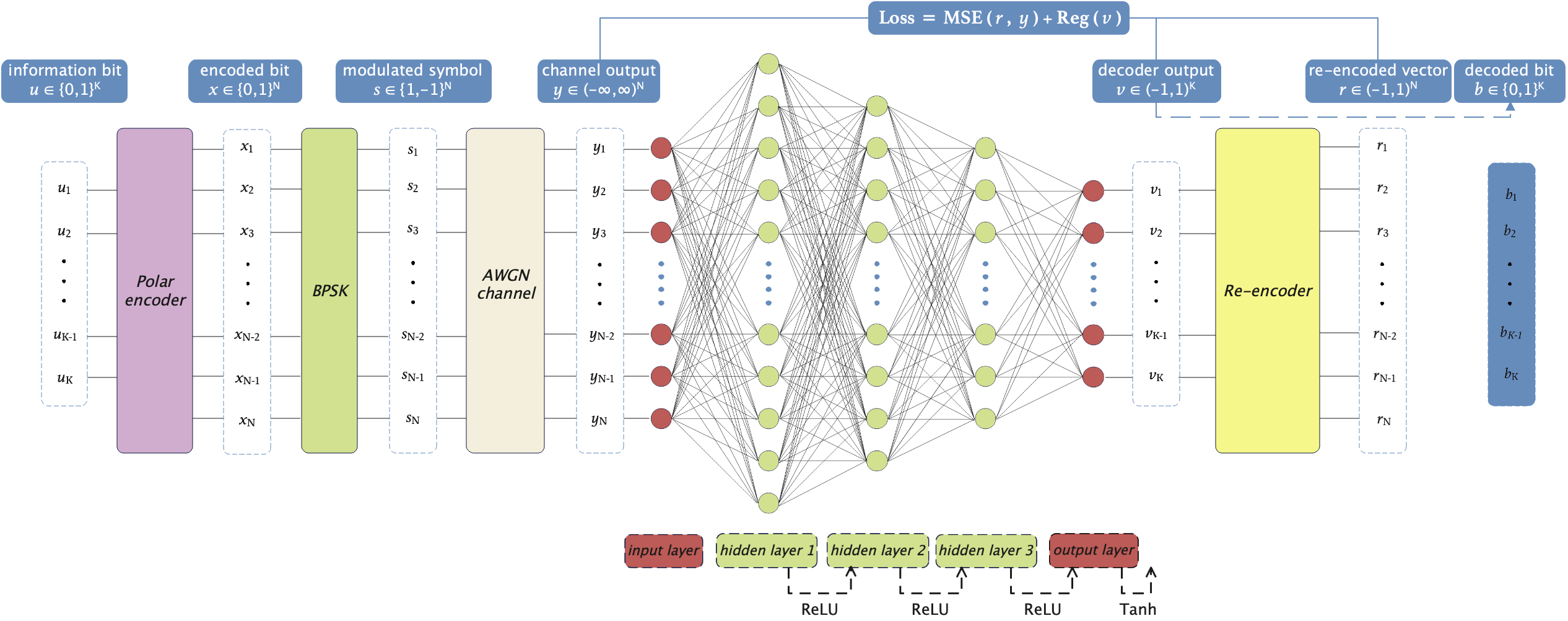}
    \caption{Coding system design and NN Architecture}
    \label{fig:NND}
\end{figure*}
In this paper, we consider the binary codes transmitted with binary phase shift keying (BPSK) modulation.
The coding system design is shown in Fig. \ref{fig:NND}. The information bit vector $u$ of length $K$ is transformed to the encoded bit vector $x$ of length $N$ by polar encoder.  Then, the BPSK modulated symbol vector $s$ of length $N$ is transmitted through AWGN channels. Subsequently, the noisy word vector $y$ of length $N$ output by AWGN channels is fed into the NND, and the NND outputs a vector $v$ of length $K$ containing values $v_i \in (-1,1), 1\leq i\leq K$. The decoded bit vector $b$ is straightforwardly given by
\begin{eqnarray}
		b_{i}= \left\{
		\begin{aligned}
			&0,\quad  \mathrm{if} \ v_{i} \geq 0,\\
			&1,\quad  \mathrm{if} \ v_{i}\  \textless \  0,
		\end{aligned}
		\right.
\end{eqnarray}
where $b_{i}$ and $v_{i}$ are the $i$-th decoded bit and decoded value, respectively.

For training NND without the ground truth (information bit vector), we involve a self-supervised module called re-encoder which operates on the NND output $v$ to obtain the valid center $r$ (re-encoded vector) of a Hamming ball containing the channel output vector $y$. This is the key innovation of our work.  

\subsection{Training NND with Re-encoder}
\begin{figure}[h]
    \centering
    \includegraphics[width=0.5\textwidth]{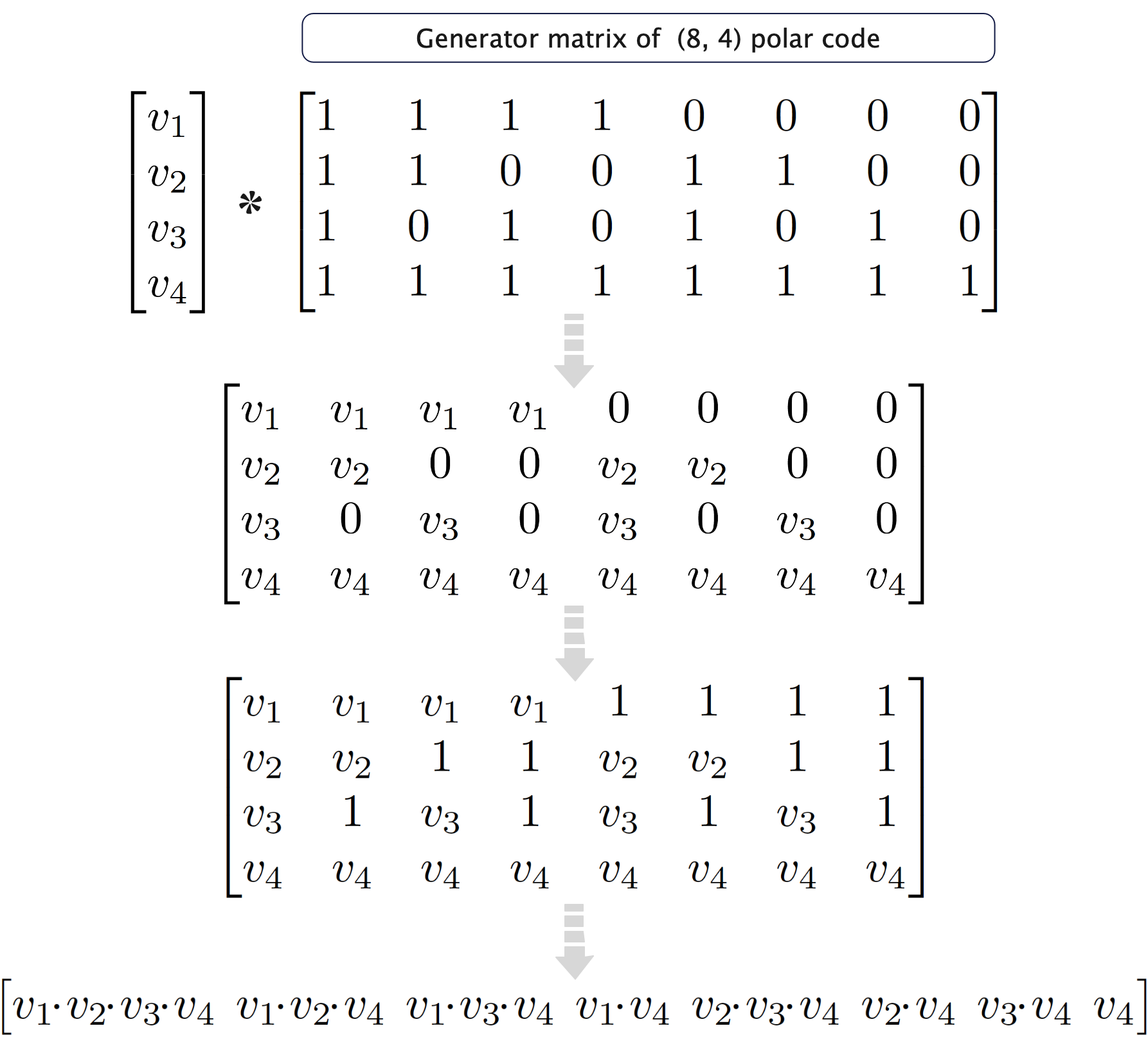}
    \caption{Illustration of how the re-encoder operates}
    \label{reencoder}
\end{figure}

The re-encoding process can be regarded as the polar encoder operates again on the NND output $v$. Nevertheless, it cannot be achieved straightforwardly by matrix multiplication, because the polar encoding is a linear transformation on $\mathbb{Z}_2$ performing on binary vectors, while $v$ is a real number vector. A plausible alternative is to binarize $v$ into the decoded bit vector $b$. However, binarization blocks the back propagation of the gradient since it is not a differentiable computation. Therefore, we define a differentiable algebraic operation on $\mathbb{R}$ to realize this encoding transformation. To be more specific, firstly, the NND output $v$ takes the Hadamard product \cite{b15} with the columns of the generator matrix to form a new $K \times N$ matrix. Then the zero entries of the the new matrix are replaced with one. Finally, the re-encoder output $r$ is obtained by columnwise multiplying the entries of the modified matrix. Fig. \ref{reencoder} shows a toy example illustrating this operation. Since $v_{i} \in (-1, 1)$, and the sign of each $v_{i}$ determines the corresponding decoded bit $b_{i}$ as described in (5), $r_{i}$ is also in $(-1, 1)$ with the sign of it representing the corresponding re-encoded bit following the same rule as (5). For instance, in the binary case, if $v=(0, 0, 1, 1)$, then the re-encoded bit vector is $r=(0, 1, 0, 1, 0, 1, 0, 1)$. Correspondingly, in the real number domain, if the signs of $v$ are $(+, +, -, -)$, then the signs of $r$ are $(+, -, +, -, +, -, +, -)$ by the operation described in Fig. \ref{reencoder}, which can give the same re-encoded bit vector. In general, the parity of the input and output of the re-encoder is represented by positive and negative signs, which in the binary case is represented by 0 and 1.

The proposed calculation rules ensures the gradient propagation of real value functions, working as the loss, along the NND and re-encoder. This design paves the foundation for training the NND by self-supervision. The loss function we use here is based on the mean squared error (MSE) of the re-encoded vector $r$ and the channel output $y$. We also involve a regularization term about $v$ to enhance the computing feasibility:
    \begin{equation}
     \mathrm{Loss}=\frac{1}{N}\sum_{i}^{N}(r_{i}-y_{i})^{2} + w\cdot \frac{1}{K}\sum_{j}^{K}||\frac{1}{v_j}||,
    \end{equation}
    where the regularization term prevents the NND output $v$ from trivial result $0$-vector and $w$ is a learn-able weight.

 By updating the weights of NN that minimize the loss function using gradient descent optimization method \cite{b16}
and the backpropagation algorithm \cite{b13}, the NND is trained to produce a vector, corresponding to a potential recovered information bits vector, of which the re-encoded
vector exhibits the smallest distance with the input of NND. This method guarantees the effective recovery of the information vector, since the correct decoded bit vector should generate a codeword that has the smallest Hamming distance with the noisy word (compared to the Hamming distances between the noisy word and all other valid codewords), when the power of the channel noise maintains controllable and the disjoint condition of the Hamming balls is satisfied as discussed in \ref{subset:MHamming}.

\section{Computer Simulations}
In our experiments, PyTorch, a widely-used deep learning framework, is employed to train the model depicted in Fig. \ref{fig:NND}. We use a 16GB MacBook M1 Pro to conduct the experiments. More detailed settings are listed as follows:\\
\begin{itemize}

    \item The NN contains an input layer with 16 nodes, an output layer with 8 nodes, and three hidden layers with 128, 64 and 32 nodes, respectively, for (16, 8) polar code.
    \item Rectified linear unit (ReLU) activation functions are used after the three hidden layers, while a tangent hyperbolic (Tanh) activation function is used after the last layer. 

     \item We adopt the Adam \cite{b17} as the optimizer and use the OneCycleLR scheduler to promote the performance and effectiveness of the training.  
     
\end{itemize}
Moreover, we want to highlight two differences with the settings in the conventional scheme \cite{b6} during the training process, which contribute to the applicability of our scheme:
\begin{itemize}
    \item We use randomly generated information bit vector, rather than the entire training set (all the different information bit vectors). For example, for (16, 8) polar code, in each training epoch, the entire training set ($2^{8}$ different information bit vectors) is passed to the system in the conventional scheme. Contrarily, the $2^{8}$ information bit vectors are randomly generated in the proposed scheme.
    \item The noise is generated with a random $E_{b}/N_{0}$ ranging from 0 dB to 10 dB in the proposed scheme, rather than a predetermined $E_{b}/N_{0}$, which is set as 1 dB in the conventional scheme. 
\end{itemize}
To ensure reproducibility of our work, we have made the source code publicly available: https://github.com/Huiying-Song/Unsupervised$\_$DL$\_$Polar$\_$Decoder.

\subsection{Performance on Very Short Packets}
\begin{figure}[h]
    \centering
    \includegraphics[width=0.47\textwidth]{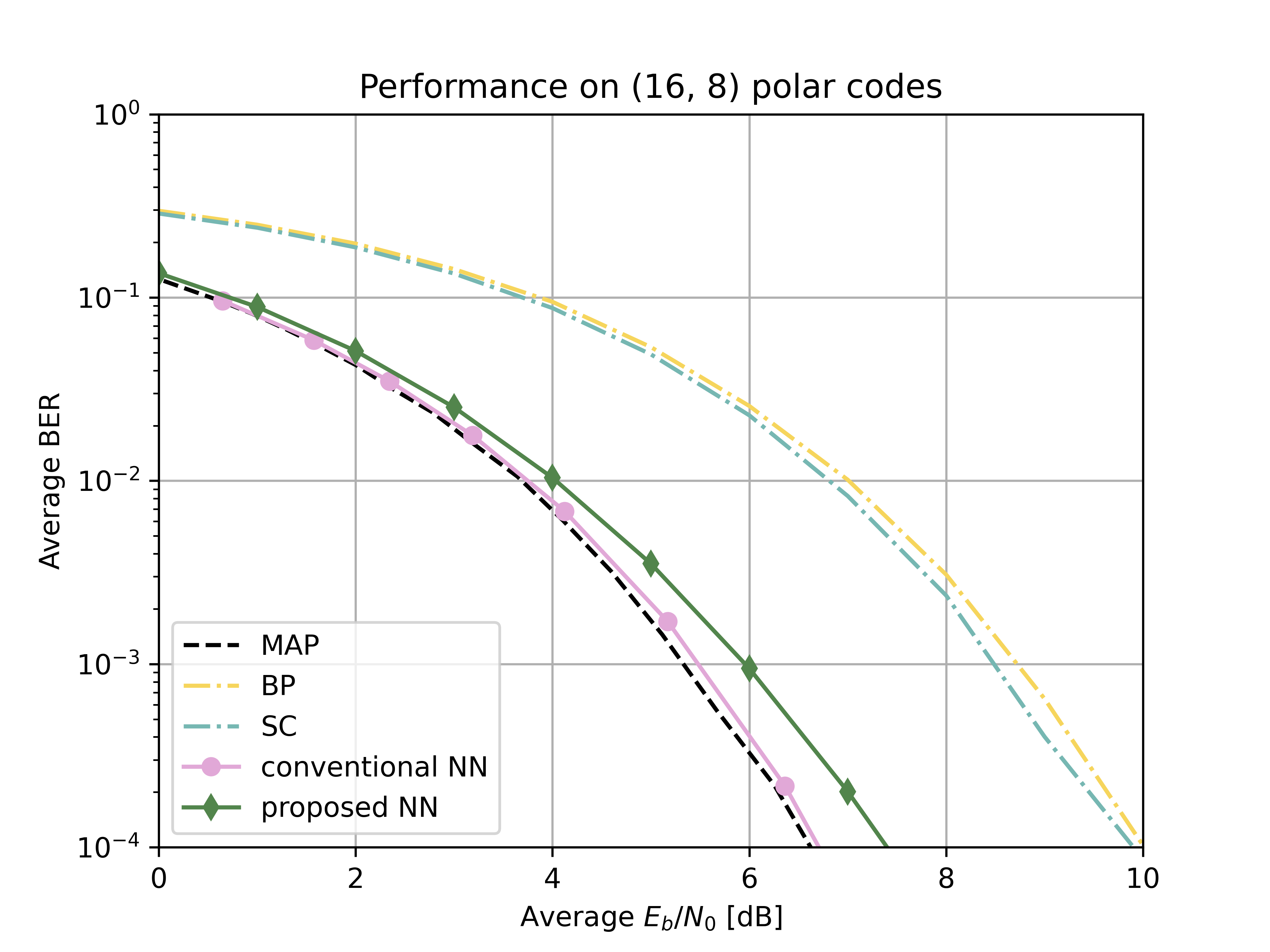}
    \caption{BER comparisons (the conventional and proposed NNs are trained with $2^{18}$ and $2^{15}$ epochs, respectively.)}
    \label{BER_performance}
\end{figure}

\begin{figure}[h]
    \centering
    \includegraphics[width=0.47\textwidth]{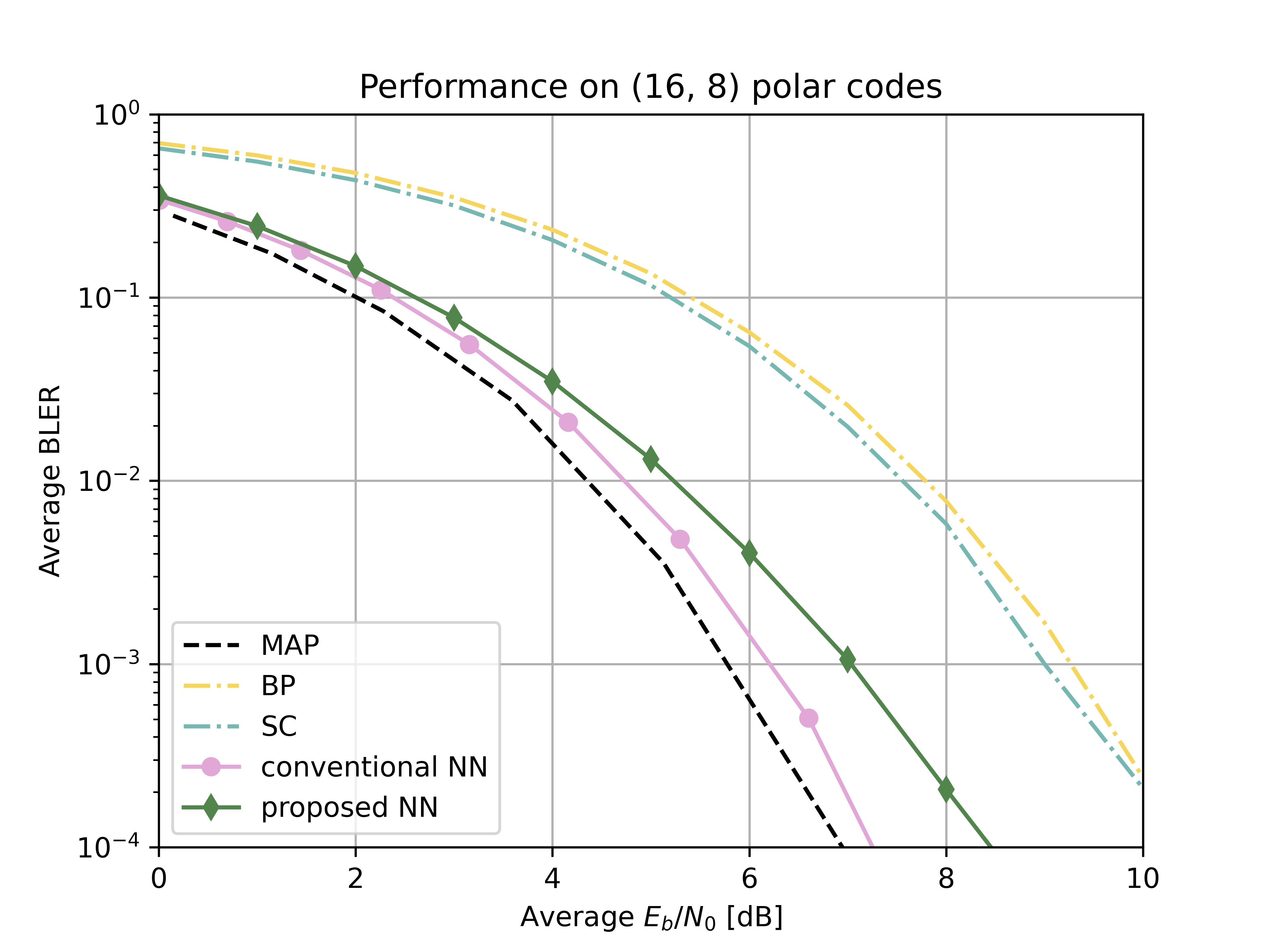}
    \caption{BLER comparisons (the conventional and proposed NNs are trained with $2^{18}$ and $2^{15}$ epochs, respectively.)}
    \label{BLER_performance}
\end{figure}

The BER and BLER performance comparisons are shown in Fig. \ref{BER_performance} and Fig. \ref{BLER_performance}. We can observe that both of the NNDs (conventional and proposed) exhibit the capability of approaching the MAP decoder, while outperform the SC and BP decoders by a significant margin.

\subsection{Generalization Ability}
\begin{figure}[h]
    \centering
    \includegraphics[width=0.47\textwidth]{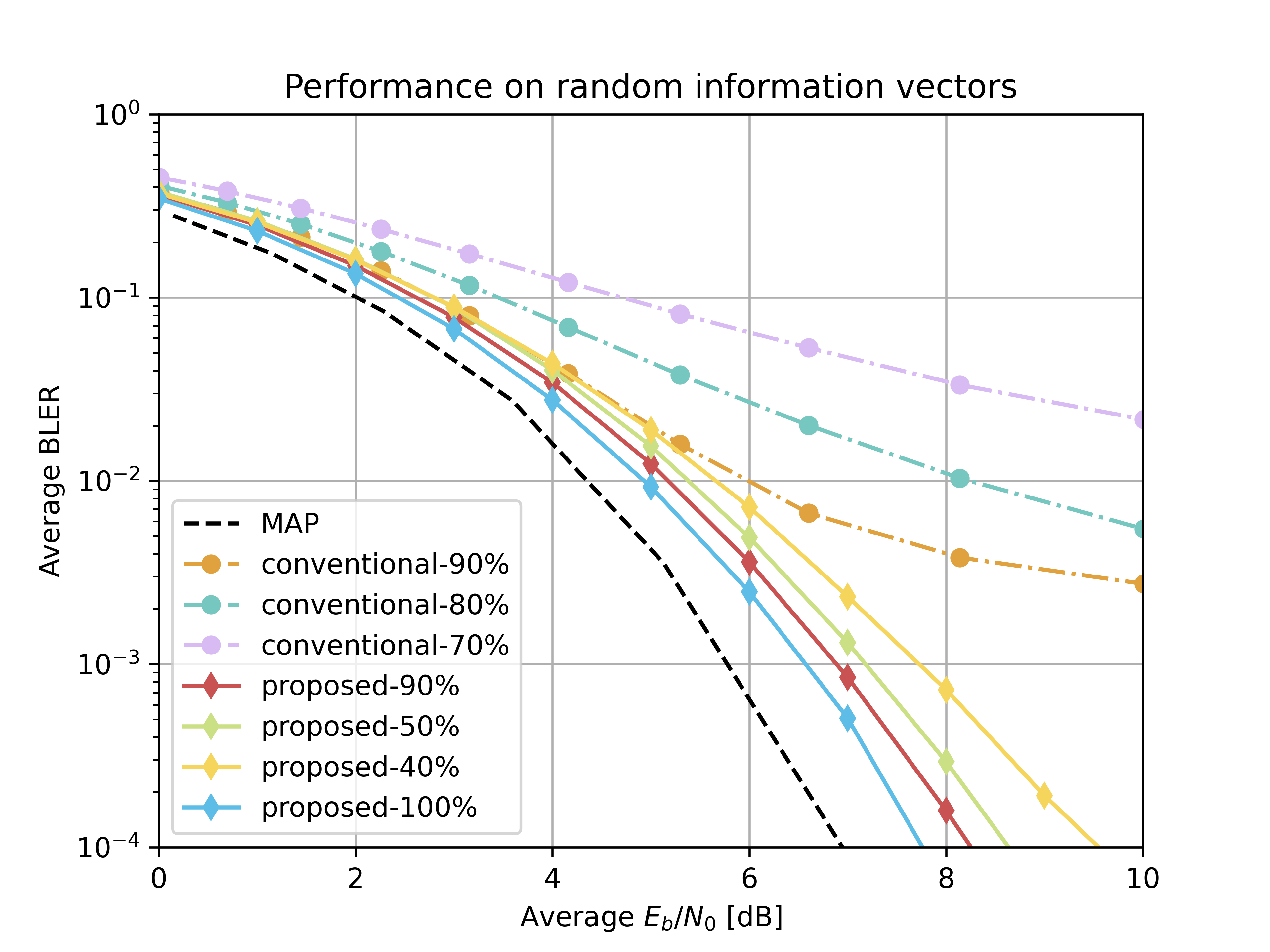}
    \caption{BLER comparisons on random information vectors (the conventional NN is trained with $2^{18}$ epochs, while the proposed NN is trained with $2^{14}$ epochs for $100 \%$ and $90 \%$, and $2^{15}$ epochs for $50 \%$ and $40 \%$, respectively.)}
    \label{generalization_random}
\end{figure}

\begin{figure}[h]
    \centering
    \includegraphics[width=0.47\textwidth]{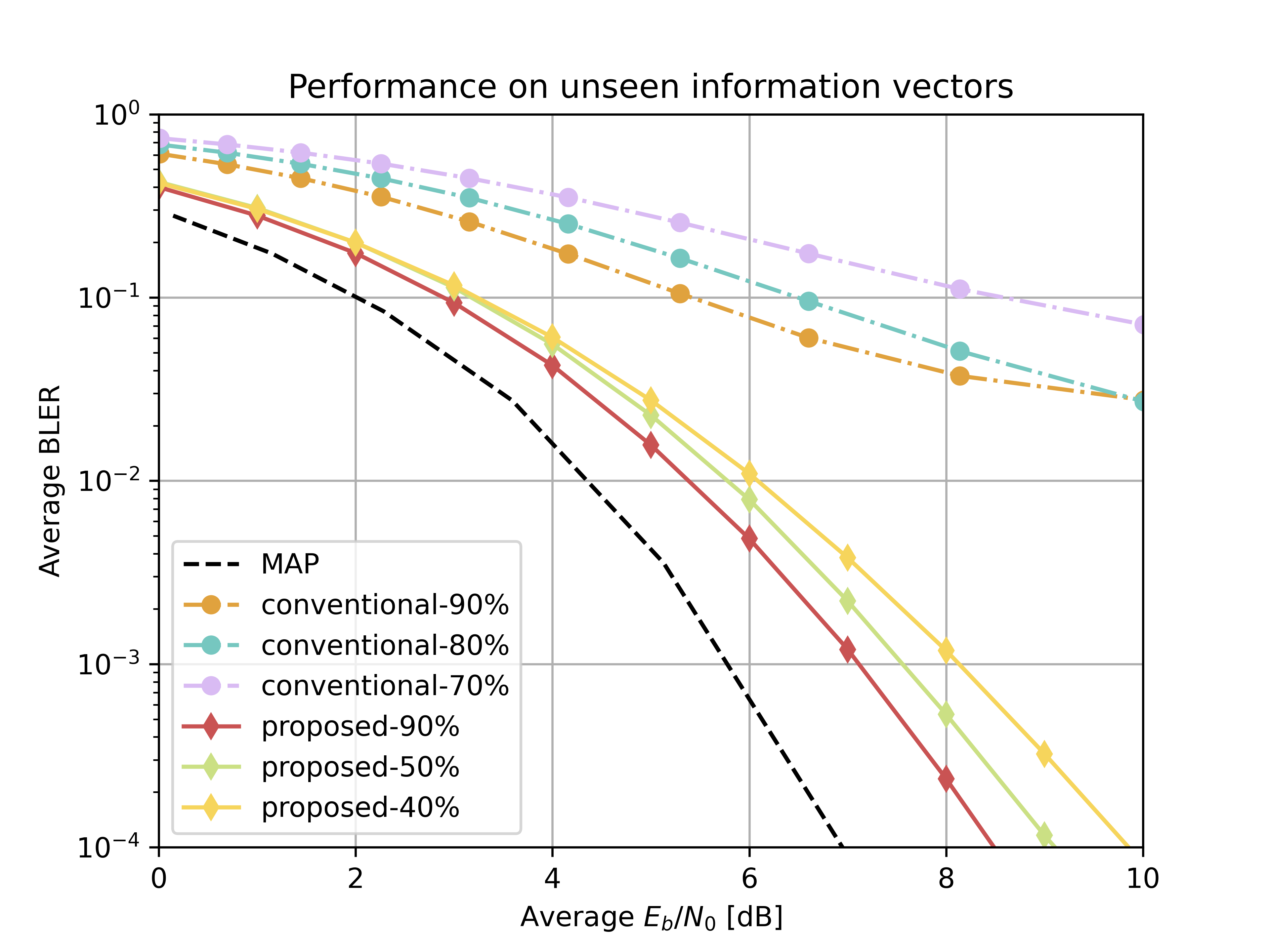}
    \caption{BLER comparisons on random information vectors (the conventional NN is trained with $2^{18}$ epochs, while the proposed NN is trained with $2^{14}$ epochs for $90 \%$ and $2^{15}$ epochs for $50 \%$ and $40 \%$, respectively.)}
    \label{generalization_unseen}
\end{figure}

To assess the generalization capability of the proposed NND, we conduct experiments in which we randomly select $p \%$ of the information bit vectors from the training set to train the NN. Subsequently, we evaluate the BLER performances of the decoder on two separate sets: (1) randomly generated information bit vectors and (2) the remaining portion of the training set. In the experiments, we test the performances of the proposed NN trained with training percentages $100 \%$, $90 \%$, $50 \%$ and $40 \%$, and compare with those of the conventional NN with training percentages $90 \%$, $80 \%$ and $70 \%$.\\
\indent The results presented in Fig. \ref{generalization_random} and Fig. \ref{generalization_unseen} demonstrate the significant improvements achieved by the proposed NND in both testing scenarios. Remarkably, the performance of the proposed NND trained with $40 \%$ of the training set surpasses that of the conventional decoder trained with $90 \%$. Additionally, it is evident that the performance gap between the two testing scenarios of the proposed scheme is much smaller compared to that of the conventional scheme, which emphasizes the effectiveness of the proposed scheme in generalizing to unseen data, showcasing its potential for utilization in communication systems with limited training time and data. Furthermore, a significant performance gap is observed between the conventional NND trained with training percentages of $90 \%$ and $100 \%$ (as the purple line depicted in Fig. \ref{BLER_performance}). In contrast, this performance gap of the proposed scheme is minor. Hence, this stark contrast suggests the possibility that the conventional NND may predominantly rely on memorizing the training set rather than actually learning the underlying decoding principle, and further provides evidence for the ability of the proposed NND to truly learn the decoding algorithm.

\section{Conclusion and Future Works}
We have proposed a self-supervised deep learning-based decoding scheme for polar codes, enabling efficient one-shot decoding. Unlike the conventional scheme that relies on using the information bit vectors as labels for training the NN, our approach leverages the generator matrix of polar codes to train the NN as a bounded distance decoder by the utilization of a re-encoder. The advantage of not requiring predefined labels and predetermined SNRs makes the proposed NND more applicable in communication systems. Through computer simulations, it has been demonstrated that the proposed scheme achieves great BER and BLER performances for very short packets. Furthermore, the generalization capability of the proposed NND greatly outperforms that of the conventional one, which contributes to the applicability in the scenarios where time and data availability are constrained.\\
\indent Scaling to practical blocklengths and adapting to more complex communication scenarios, such as the orthogonal frequency division multiplexing (OFDM) \cite{b18} transmission with quadrature phase shift keying (QPSK) modulation \cite{b19} over frequency selective fading channels, are the future works. In addition, this decoding approach can be applied for any linear codes with a generator matrix.

\section*{Acknowledgment}
We would like to thank Kiyomichi Araki for his insightful discussions and valuable comments on the paper, which is instrumental in shaping the direction of the work.

\vspace{12pt}


\begin{thebibliography}{00}
\bibitem{b1} I. Tal and A. Vardy, “List Decoding of Polar Codes,” \textit{IEEE Transactions on Information Theory}, vol. 61, no. 5, pp. 2213-2226, May 2015, doi: 10.1109/TIT.2015.2410251.

\bibitem{b2} E. Arıkan, “Polar codes: A pipelined implementation,” in \textit{4th International Symposium on Broadband Communication (ISBC 2010)}, July 2010.

\bibitem{b3} E. Nachmani, Y. Be'ery and D. Burshtein, “Learning to decode linear codes using deep learning,” in \textit{2016 54th Annual Allerton Conference on Communication, Control, and Computing (Allerton)}, Monticello, IL, USA, 2016, pp. 341-346, doi: 10.1109/ALLERTON.2016.7852251.

\bibitem{b4} W. Xu, Z. Wu, Y. -L. Ueng, X. You and C. Zhang, “Improved polar decoder based on deep learning,” in \textit{2017 IEEE International Workshop on Signal Processing Systems (SiPS)}, Lorient, France, 2017, pp. 1-6, doi: 10.1109/SiPS.2017.8109997.

\bibitem{b5} B. Yuan and K. K. Parhi, “Architecture optimizations for BP polar decoders,” in \textit{2013 IEEE International Conference on Acoustics, Speech and Signal Processing (ICASSP)}, Vancouver, BC, Canada, 2013, pp. 2654-2658, doi: 10.1109/ICASSP.2013.6638137.

\bibitem{b6} T. Gruber, S. Cammerer, J. Hoydis and S. t. Brink, “On deep learning-based channel decoding,” in \textit{2017 51st Annual Conference on Information Sciences and Systems (CISS)}, Baltimore, MD, USA, 2017, pp. 1-6, doi: 10.1109/CISS.2017.7926071.

\bibitem{b7} N. Doan, S. Ali Hashemi and W. J. Gross, “Neural Successive Cancellation Decoding of Polar Codes,” in \textit{2018 IEEE 19th International Workshop on Signal Processing Advances in Wireless Communications (SPAWC)}, Kalamata, Greece, 2018, pp. 1-5, doi: 10.1109/SPAWC.2018.8445986.

\bibitem{b8} S. Cammerer, T. Gruber, J. Hoydis and S. ten Brink, “Scaling Deep Learning-Based Decoding of Polar Codes via Partitioning,” in \textit{GLOBECOM 2017 - 2017 IEEE Global Communications Conference}, Singapore, 2017, pp. 1-6, doi: 10.1109/GLOCOM.2017.8254811.

\bibitem{b9} V. B. Misic and J. Misic, \textit{Machine-to-Machine Communications: Architectures, Technology, Standards, and Applications}. CRC Press, 2017.

\bibitem{b10} G. Durisi, T. Koch and P. Popovski, “Toward Massive, Ultrareliable, and Low-Latency Wireless Communication With Short Packets,” in \textit{Proceedings of the IEEE}, vol. 104, no. 9, pp. 1711-1726, Sept. 2016, doi: 10.1109/JPROC.2016.2537298.

\bibitem{b11} E. Arikan, “Channel Polarization: A Method for Constructing Capacity-Achieving Codes for Symmetric Binary-Input Memoryless Channels,” in \textit{IEEE Transactions on Information Theory}, vol. 55, no. 7, pp. 3051-3073, July 2009, doi: 10.1109/TIT.2009.2021379.


\bibitem{b12} R. Horn and C. R. Johnson, \textit{Matrix Abalysis}. Cambridge University Press, 1985.

\bibitem{b13} D. E. Rumelhart, G. E. Hinton, and R. J. Williams, “Parallel distributed processing: Explorations in the microstructure of cognition, vol. 1.” Cambridge, MA, USA: MIT Press, 1986, pp. 318–362.


\bibitem{b14} T. Richardson and R. Urbanke, \textit{Modern Coding Theory}. Cambridge University Press, 2008.

\bibitem{b15} Hadamard, “J. Essai sur l'étude des fonctions données par leur développement de Taylor,” \textit{Journal de Mathématiques Pures et Appliquées, Serie 4}, Volume 8 (1892), pp. 101-186. 

\bibitem{b16} G. E. Hinton, S. Osindero, and Y.-W. Teh, “A fast learning algorithm for
deep belief nets,” \textit{Neural Computation}, vol. 18, no. 7, pp. 1527–1554,
July 2006.

\bibitem{b17} D. P. Kingma and J. Ba, “Adam: A method for stochastic optimization,” in \textit{International Conference on Learning Representations (ICLR)}, 2015. [Online]. Available: https://arxiv.org/abs/1412.6980




\bibitem{b18} J. Terry and J. Heiskala, \textit{OFDM Wireless LANs: A Theoretical and Practical Guide}. Sams, 2001.

\bibitem{b19} J. G. Proakis and M. Salehi, \textit{Digital Communications}. McGraw-Hill, 2008.
\end{thebibliography}
\end{document}